
\frontpagetrue
\title{A naive matrix-model approach to two-dimensional quantum gravity
coupled to matter of arbitrary central charge}
\author{ Edouard Br\'ezin}
\address{ Laboratoire de Physique Th\'eorique de l'Ecole Normale
Sup\'erieiure, 24 rue Lhomond 75231, Paris Cedex 05, France
\foot{Unit\'e prope du Centre National de la Recherche Scientifique,
associ\'ee ${\grave a}$ l'Ecole Normale Sup\'erieure et ${\grave a}$
l'Universit\'e Paris-Sud} }
\author{ Shinobu Hikami }
\address{ Department of Pure and Applied Sciences, University of Tokyo,
Meguro-ku, Komaba, 3-8-1, Tokyo 153, Japan}

\abstract{
In the usual matrix-model approach to random discretized two-dimensional
manifolds, one introduces n Ising spins on each cell, i.e. a discrete
version of 2D quantum gravity coupled to matter with a central charge
n/2.  The matrix-model consists then of an integral over $2^n$ matrices,
which we are unable to solve for $n>1$.  However for a fixed genus we
can expand in the cosmological constant g for arbitrary values of n,
and a simple minded analysis of the series yields for n=0,1 and 2 the
expected results for the exponent $\gamma_{string}$ with an amazing
precision given the small number of terms that we considered.  We then
proceed to larger values of n.  Simple tests of universality are
successfully applied; for instance we obtain the same exponents for
n=3 or for one Ising model coupled to a one-dimensional target space.
The calculations are easily extended to q-states Potts models, through
an integration over $q^n$ matrices.  We see no sign of the tachyonic
instability of the theory, but we have only considered genus zero at
this stage.
\break
LPTENS 92/10                 \hskip 9cm March 1992}

The exactly soluble matrix-models representations of matter coupled to
2D gravity are not very numerous.  With one matrix the models are
solvable\ref{1} but the central charge is restricted to the (2,2k-1)
models and thus $c \leq 0$.  With several matrices the existing
technologies are limited to an open chain of matrices coupled to their
nearest neighbours\ref{2} which describe matters with c at most equal
to one (in the limit of an infinite chain).  It is vesry easy to
formulate with matrix models matter fields with $c>1$, but one does
not seem at present to be able to solve these models analytically.
Presumably it is interesting to examine these models numerically for
finite dimensional matrices (although it would involve an analytic
continuation in the coupling constant in  most cases), but we have
simply used here straightforward perturbative series in the
cosmological constant at fixed genus.  The critical cosmological
constant and the exponents are related to the behaviour of the
coefficients of the series for increasing order.
\par
The model of n Ising spins, coupled to gravity, is deduced as for
one-Ising models\ref{3}, from the discretized Polyakov representation\ref{4}
of closed bosonic string theories.  At each cellcenter of the
discretized manifold we have n Ising variables
$(\sigma_{1},\sigma_{2},\dots ,\sigma_{n})$
and adjacent cells of the triangulation are coupled with a Boltzmann
weight:
$\exp \beta (\sigma_{1}\sigma '_{1}+\sigma_{2}\sigma '_{2}+\dots
+\sigma_{n}\sigma '_{n}-n)$.
For simplicity we have chosen the same coupling constants for all the
spins, so that they all become critical together. In the matrix-model
representation of the triangulation we have to introduce a different
matrix for every spin-configulation of the cell, and thus $2^n$
Hermitian $N \times N$ matrices $\Phi_{\tau}$, in which $\tau$
stands for the n-plet
$(\sigma_{1},\sigma_{2},\dots ,\sigma_{n})$.
The matrix model is thus simply given by the action

$$
S = N \Tr \biggl[ {1 \over 2} \sum_{\tau ,\tau '}(\Phi_{\tau}\Delta_{\tau ,\tau
'}
\Phi_{\tau '})+ g\sum_{\tau}\Phi_{\tau}^{4} \biggr]  \eqno{(1)}
$$

\noindent
in which $\Delta_{\tau ,\tau '}$ is such that the propagator
$(\Delta^{-1})_{\tau ,\tau '}$ is proportional to the Boltzmann
weight $\exp\beta\tau \cdot \tau '$ where
$\tau \cdot\tau '=\sigma_{1}\sigma '_{1}+\sigma_{2}\sigma '_{2}\dots
+\sigma_{n}\sigma '_{n}$.
It is straightforward to verify that

$$
if (\Delta^{-1})_{\tau ,\tau '} =\exp \beta\tau\cdot\tau ',
then (\Delta )_{\tau ,\tau '} =
{1 \over {(2 \sinh 2\beta )^{n}}}\hat \tau \hat \tau ' \exp\beta\tau\cdot\tau '
\eqno{(2)}
$$

\noindent
in which $\hat \tau =\sigma_{1}\sigma_{2}\dots\sigma_{n}$.

One can also introduce a one-dimensional version with Ising degrees of
freedom, with a total matter central charge
$c=1 + n/2$.
The model is a one-dimensional path-integral over $2^n$ matrices, with
an action

$$
S = N \int dt Tr \biggl[{1 \over 2} \sum_{\tau ,\tau '}\Delta_{\tau ,\tau '}
(\dot \Phi_{\tau}\dot \Phi_{\tau '}+m^2 \Phi_{\tau}\Phi_{\tau '})
+g\sum_{\tau} \Phi_{\tau}^{4} \biggr]. \eqno{(3)}
$$

The expansion in powers of $g$ and $1/N^2$ yields Feynman diagrams
of fixed genus\ref{5}.  For a given diagram, which is simply the dual
of a fixed discretized 2D manifold, the n Ising spins are decoupled:
the Ising partition function for the spins is thus simply the n-th
power of a single Ising partition function associated to the diagram.
Therefore the diagrammatic rule is the following:  we attribute to
each vertex of the diagram a spin index $\sigma$; to each propagator
of the diagram connecting two vertices with spins $\sigma$ and
$\sigma '$ we associate the factor $a^{(1-\sigma\sigma ')/2}$,
in which

$$
a = e^{-2\beta}  \eqno{(4)}
$$

We sum over all spins and raise this sum to the power $n$.  This rule
applies to zero and one-dimensional Feynman diagrams: the n-Ising
partition function multiplies the contribution of the diagram in the
absence of Ising spins, i.e. the contribution of pure gravity for
zero-dimensional matrix models, or of $c=1$ matter coupled to gravity
for one-dimensional matrix models.  Pure gravity, or the $c=1$ model,
are thus recovered in the $n=0$ limit.

These rules are easily generalized to q-state Potts matter fields
coupled to gravity.  For one Potts "spin" one introduces q matrices
instead of two for Ising spins.  Each vertex of the Feynman diagram
has a "colour" $\sigma$, $(\sigma =1,2,\dots ,q)$, and a propagator
connecting two vertices with colours $\sigma$ and $\sigma '$ yields
an additional factor

$$
a^{(1-\delta_{\sigma ,\sigma '})/2}.  \eqno{(5)}
$$

One then computes the Potts partition function of the Feynman diagrams;
again n Potts spins per plaquette, one raises the partition function
of each diagram to the power n.  We have evaluated up to order
$g^6$  the free energy in the genus zero case.  The expression is very
long and we reproduce here simply the free energy $F(g)$ for n
q-states Potts model up to order $g^4$:

$$
\eqalign{&{F \over {q^{n}}}
          =2g-g^2 \biggl( 16[1+(q-1)a^2]^n +2[1+(q-1)a^4]^n \biggr)\cr
          &+g^3\biggl( 128[1+(q-1)a^2]^{2n}+{256\over {3}}
          [1+3(q-1)a^2 +(q-1)(q-2)a^3]^n \cr
          &+{32\over {3}} [1+3(q-1)a^4 +(q-1)(q-2)a^6 ]^n \cr
          &+64[1+(q-1)a^2 +2(q-1)a^4 +(q-1)(q-2)a^5 ]^n \biggr) \cr
          &+g^4 \biggl( 1024[1+(q-1)a^2 ]^{3n} \cr
          &+2048[1+4(q-1)a^2 +(q-1)(q-2)a^3 +3(q-1)^2 a^4 +(q-1)^2
          (q-2)a^5 ]^n \cr
          &+512[1+6(q-1)a^2 +4(q-1)(q-2)a^3 +(q-1)(q^2 -3q+3)a^4 ]^n \cr
          &+64[1+(q-1)a^2 +4(q-1)a^4 +2(q-1)(q-2)a^5 +(q-1)a^6 \cr
          &+2(q-1)(q-2)a^7 +(q-1)(q^2 -3q+3)a^8 ]^n \cr
          &+32[1+6(q-1)a^4 +4(q-1)(q-2)a^6 +(q-1)(q^2 -3q+3)a^8 ]^n }
$$
$$
\eqalign{
&+512[1+(q-1)a^2 +5(q-1)a^4 +2(q-1)(q-2)a^5 +(q-1)(3q-5)a^6 \cr
&+(q-1)(q-2)^2 a^7 ]^n \cr
&+768[1+2(q-1)a^2 +(q-1)(q+3)a^4 +4(q-1)(q-2)a^5 \cr
&+(q-1)(q-2)^2 a^6 ]^n \cr
&+512[1+3(q-1)a^2 +(q-1)(q-2)a^3 +3(q-1)a^4 +3(q-1)(q-2)a^5 \cr
&+(q-1)(q^2 -3q+3)a^6 ]^n \cr
&+512[1+2(q-1)a^2 +(q-1)(q+1)a^4 +(q-1)(q-2)a^5 +2(q-1)^2 a^6 \cr
&+(q-1)^2 (q-2)a^7 ]^n \cr
&+64[1+5(q-1)a^4 +2(q-1)^2 a^6 +4(q-1)(q-2)a^7 \cr
&+(q-1)(q-2)(q-3)a^8 ]^n \biggr) } \eqno{(6)}
$$

Pure gravity is recovered from several limits:a=0 or a=1 or n=0. The
limit q goes to one corresponds to percolation on a random lattice,
and q=2 to n Ising models. The expansion in powers of $g$,
for genus zero,

$$
{F \over {q^{n}}} = \sum_k C_k g^k  \eqno{(7)}
$$

behaves for large orders as

$$
C_k \simeq A^k k^{-3+\gamma_{string}}. \eqno{(8)}
$$

The amplitude A is the inverse of the critical cosmological
constant, whereas $\gamma_{string}$, the string susceptibility
exponent is universal.
\par
  In this letter, we use several methods of analysis to extract
$A$ and $\gamma_{string}$ from the expansion.  The ratio of the
coefficients $f_k =C_k /C_{k-1}$ is given by Pad\'e approximation,

$$
f_k = A \biggl({ {1+{{b_1}\over {k}}+{{b_2}\over{k^2}}+\dots}\over
{1+{{c_1}\over {k}}+{{c_2}\over {k^2}}+\dots}} \biggr) \eqno{(9)}
$$
\noindent
in which $b_1 - c_1 =-3+\gamma_{st}$ for $k \rightarrow \infty$.  We
first see the pure gravity case is obtained for a=0, a=1 or n=0.
{}From the exact solution of the planar pure gravity\ref{5b},
one finds that by the [2,1] Pad\'e with $b_{1}=-3/2$,$b_{2}=1/2$,
$c_{1}=2$, $\vert A \vert=48$ and $\gamma_{string}=-1/2$. The Pad\'e
coefficients $b_{m}$ and $c_{m}$ are determined by the values of
$C_{k}$ in (7).  The second derivative of
$\ln A$ with respect to $a$ is
proportional to the specific heat of n q-state
Potts models on a random surface.  This Pad\'e approximation gives
simultaneously this specific heat and the value of $\gamma_{st}$
as a function of the parameter $a$.  There is a critical value of $a$,
at which, had we calculated exactly, we would see a discontinuous
change from the pure gravity criticality $\gamma_{string}=-1/2$,
to the actual matter coupled to gravity exponent. In our numerical
work, we find instead a bump for $\gamma_{string}$ as a function
of $a$, and we need an extrapolation procedure to go to the
large k limit.
\par
For n=1 Ising case, we obtain $\gamma_{st}=-0.2813$ for [2,1]
Pad\'e of order $g^5$ and $\gamma_{st}=-0.3328$ for the order $g^6$
series.  They are very close to the exact value of -1/3.  The peak of
the specific heat, which is proportional to $d^2 \ln A/d^2 a$,
appears at $a=0.25$, which is the exact value\ref{3}.
  Although the sudden change of $\gamma_{st}$ from -1/2 to
-1/3 is expected at the critical point $a=0.25$, our Pad\'e analysis
shows a gradual crossover to the new fixed point.
  The width of the
peak of $\gamma_{st}$ is indeed reduced by a factor 2/3 from the
result of order $g^5$ to $g^6$.  We then use [1,1] Pad\'e for
observing the successive narrowing of the width of the peak and we
extrapolate the value of $\gamma_{st}$ in the large $k$ limit.  This
[1,1] Pad\'e is given by
$$
f_k = A(1- {1 \over k} )\biggl({ {1+{b{_1}\over k}}\over
{1+{c_{1}\over k}}}
\biggr) \eqno{(10)}
$$
\noindent
which is consistent with the exact result of the pure gravity case,
with $b_1 =-1/2$, $c_1 =2$. The value of $\gamma_{st}=2+b_1 -c_1$
shows clearly the narrowing from $O(g^5 )$ to $O(g^6 )$ in Fig.1 by
this [1,1] Pad\'e analysis, which may give a
reasonable extrapolation
to the large order limit.  The n=1, q=4 Potts model
 has a maximum peak of $\gamma_{st}=-0.032$ at
$a=0.26$ in [2,1] Pad\'e analysis of order $g^5$.  Let us recall that
the predicted
value of $\gamma_{st}$ of q-state Potts model with 2D gravity is\ref{3}
$$
\gamma_{st}={1 \over {1- {\pi\over {\arccos ({\sqrt {q}\over {2}})}}} }
\eqno{(11)}
$$
\noindent
for $0\leq q \leq 4$.  Thus the c=1,
$q=4$ Potts model, has $\gamma_{st}=0$, and our result is very close.

The one-dimensional version of the q states-Potts model
coupled to gravity, has an extra central charge of
one unit. The Feynman diagrams are now given by one-dimensional
integrals over the loop momenta. For n=0 we recover the free
energy of the c=1 matrix model, which is soluble\ref{5b} and
we have used it as to test our expansion and extrapolations.
Up to order seven we find a free energy

$$
F={1 \over 2}g-{17\over 16}g^2 + {75\over 16}g^3 -27.83593g^4
+195.42187g^5 -1529.4917g^6 +12924.5864g^7 . \eqno{(12)}
$$

Here we employ a ratio method for obtaining accurate values of
$\gamma_{string}$. Using the ratio $f_k$, the cosmological constant
$A_k$ and $\gamma_{string}$ of order $k$ are given by

$$
A_k =(1+k)f_k -kf_{k-1} \eqno{(13)}
$$
$$
\gamma_{string}=3 - {{k(1+k)(f_k -f_{k-1})}\over {A_k}}. \eqno{(14)}
$$

\noindent
The extrapolation to $k \rightarrow \infty$ is analyzed by the
plotting $\gamma_{string}$ as a function of $1/k$, which is
shown in Fig. 2.  This ratio method
gives amazingly accurate values for $\gamma_{string}$.  For the $d=1$ one
matrix model, it yields $\gamma_{string}=0$, which is consistent with
the exact solution.  For n=2 Ising model (q=2), we have
$\gamma_{string}=0$ by this method, where the critical value of $a$ is
0.25.  The result of $\gamma_{string}=0$ for n=2 Ising model is
consistent with the universality.  It shows that the central charge of
n-Ising models is indeed $c=n/2$.  We have also confirmed by the same
analysis that the n=1, q=4 Potts model which corresponds also to c=1,
 has $\gamma_{string}=0$.
\par
For $n$ Ising models ($n \geq 3$), there is a tendency to splitting
of the peak of $\gamma_{string}$ into  double peaks when the coupling
constant  $a$ is changed from zero to one at lower orders of
perturbation.  However, the distance of this double splitting becomes
small when one increases the order of the perturbation by Pad\'e
analysis.  We have done a careful analysis by both methods, [1,1]
Pad\'e of (10) and the ratio-extrapolation method of (14).  The
transition point $a$ appears at $a < 1/4$ for $n >3$ Ising models.
  The ratio-extrapolation method is more accurate,
therefore we use this method
to extract the value of $\gamma_{string}$ for $n \geq 3$.  The results
are shown in table 1 and Fig. 3.  The value of $\gamma_{string}$ is
small and positive.  The maximum value $\gamma_{string}=0.1$ appears at
n=5 as shown in Fig. 3.  Beyond n=5 (c=2.5), the value of
$\gamma_{string}$ decreases.  For large values of $c$, we need
higher orders.  The critical point $a$ decreases, for
example a=0.1 for n=8.  We have also analyzed n, q=4 Potts model,
which has a central charge c=n.  The result coincides with the value
of $\gamma_{string}$ obtained from n-Ising models.

It may be interesting to invesigate which type of diagram becomes
important for $a \rightarrow 0$.  We find the dominant diagrams,
which have the largest coefficients of $a^2$ from (6).  Such diagrams
, which consist of  maximum bubbles on a circle, seem
to be related to the concentration of singularities on the surface. They are
different from branched polymer diagrams.
\par
In conclusion, we have obtained  new results for the string susceptibility for
$c >1$ by considering n-Potts models on a random surface.  This
result seems to be universal since it coincides with other models,
d=1 n-Ising model or n q=4 Potts model.  We have used  perturbation
series of relatively small order to estimate $\gamma_{string}$ in the genus
zero
case and we have obtained the value of $\gamma_{string}$ with an amazing
precision for arbitrary c.  The generalization to genus one or two is
easy in our method.  These results and also higher order
analysis of the seventh order will be reported elsewhere.
\vfill

\eject
\centerline{\bf References}

\item{1)} M.R. Douglas and S.H. Shenker, Nucl. Phys. {\bf B335}
           (1990) 635.
\item{ }  D.J. Gross and A.A. Migdal, Phys. Rev. Lett. {\bf 64}
           (1990) 127.
\item{ }  E. Br\'ezin and V.A. Kazakov, Phys. Lett. {\bf B236} (1990)
           144.
\item{2)} M.L. Mehta, Comm. Math. Phys. {\bf 79} (1981) 327.
\item{ }   S. Chada, G. Mahox and M.L. Mehta, J. Phys. A: Math. Gen.
           {\bf 14} (1981) 579.
\item{ }   E. Br\'ezin, M.R. Douglas, V.A. Kazakov and S.H. Shenker, Phys.
           Lett. {\bf B237} (1990) 43.
\item{ }   D.J. Gross and A.A. Migdal, Phys. Rev. Lett. {\bf 64}
           (1990) 717.
\item{ }   C. Crnkovic, P. Ginsparg and G. Moore, Phys. Lett.
           {\bf B237} (1990) 196.
\item{ }   M.R. Douglas, Phys. Lett. {\bf B238} (1990) 176.
\item{3)}  V.A. Kazakov, Phys. Lett. {\bf A199} (1986) 140.
\item{ }   D.V. Boulatov and V.A. Kazakov, Phys. Lett. {\bf B187}
           (1987) 379.
\item{4)}  F. David, Nucl. Phys. {\bf B257} (1985) 45.
\item{ }   V.A. Kazakov, Phys. Lett. {\bf B150} (1985) 45.
\item{ }   V.A. Ambjorn, B. Durhuus and J. Frohlich, Nucl. Phys.
           {\bf B257} (1985) 433.
\item{ 5a)} G. 't Hooft, Nucl. Phys. {\bf 75} (1974) 461.
\item{ 5b)} E. Br\'ezin, C. Itzykson, G. Parisi and J.B. Zuber,
           Comm. Math. Phys. {\bf 59} (1978) 35.
\vfill

\eject
\centerline{\bf Figure captions}
\item{Fig. 1} The string susceptibility $\gamma_{st}$ of n=1 Ising
              model.  The [1,1] Pad\'e results of the order $g^5$ and
              $g^6$ are shown by a) and b), respectively.  The
              narrowing of the width of the peak is obtained by
              increasing the order.  The peak of order $g^6$ yields
              $\gamma_{st}=-1/3$.
\item{Fig. 2} The string susceptibility $\gamma_{string}=0$ is obtained
              by the extrapolation of $k\rightarrow \infty$
              for d=1 one matrix model.
\item{Fig. 3} The string susceptibility $\gamma_{string}$ as a
              function of the central charge c is obtained by the
              ratio-extrapolation method for n-Ising models with
              c=n/2.
\vfill

\eject
\input tables
\centerline{\bf table 1,
The value of $\gamma_{string}$ for n-Ising model on the random
surface.}
\vskip 1cm
\def\H{\hfill}
\begintable
n=0   \H| n=1    \H| n=2  \H| n=3  \H| n=4  \H| n=5 \H| n=6  \H| n=7  \H| n=8
\crthick
 -0.5 \H| -0.333 \H|  0   \H| 0.02 \H| 0.06 \H| 0.1 \H| 0.07 \H| 0.06 \H|
0.05\endtable